# Rightsizing the Railway Signal Workforce: a Zero-Based Resourcing Approach Towards Asset Management

Alex Lu[1], Zhiqi Zhong[1], Thomas Barger[1], and Michael Brotzman[2]

[1] Metro-North Commuter Railroad, P.O. Box 684, Ossining, New York 10562-0684
Email: lexcie (at) gmail.com (corresponding author)

[2] Independent Consultant, 6425 Montgomery Road, Elkridge, Maryland 21075

## ABSTRACT

Classic asset management approaches begin by inventorying all infrastructure assets and then assigning maintenance tasks and resources to them. Our approach collects similar data, but by starting with current personnel assignment and describing their job responsibilities and work processes, staff resistance in a railroad infrastructure owner-operator environment is minimized. The resulting "manning model" quantitatively measures signal maintenance burden including Federally mandated tests, trouble tickets, non-FRA maintenance, overhead and vacation coverage, location/shift assignment, administrative process, and work curfew productivity losses. It is capable of delivering immediate results by rightsizing allocation of workforce across shifts and maintenance base locations—even before all assets are formally inventoried. Typical data from a commuter passenger railroad shows that work curfews and shift assignment constraints have significant impacts on workforce productivity. Just over half of signal maintenance employee-hours are devoted to Federally mandated tests, whilst non-FRA maintenance and repair maintenance consumes about 25% each. These indicators provide intelligence that can drive strategic management actions to improve signal maintenance cost-effectiveness. The manning model provides workload-based employee assignment by craft, location, gang, and shift for maintenance manager use, but also provides analytical basis for establishing or abolishing positions in the budgeting process. Comparing model results with current employee payroll provides a measure of how much staffing stress the maintenance organization is under, which can help measure whether the current overtime usage is appropriate. The asset and maintenance task inventories collected in this process can also feed normal asset management processes to assess replacement cycles, asset failure risk, and to inform strategic and investment decisions. This personnel-centric approach can help improve acceptance of asset management processes in the field by aligning it with maintenance managers' daily routine, and generates early benefits by improving the understanding of field maintenance activities.

**Key Words:** railroad signal maintenance, manpower analysis, resource review, asset management, workload modelling, shift assignment

## INTRODUCTION

Classic asset management approaches [1] have tended to begin by inventorying assets at a nauseating level of detail—an asset register—and creating database relationships that describe the hierarchical relationship between categories, assets, components, and sub-components. Required workforce is then allocated to assets based on numerous variables. This was often done with a view towards or as a part of contracting out maintenance work [2]. The logic flow is Assets—Actions—People.

However, this is not normally how infrastructure maintenance is actually thought of, and carried out, by field forces. Field infrastructure managers instead focuses on cyclical inspection and preventative maintenance tasks lists that must be completed within a prescribed timeframe, and any backlog repair tasks (prioritized list typically generated from issues reported on trouble tickets or asset condition findings recorded during field inspections); the workflow is instead People—Tasks—Assets. By adopting a personnel-centric approach towards asset management, it dramatically improves everyone's (including field forces') understanding of the criticality of this process, because it aligns the higher-level asset management framework with daily routine of dispatching employees to locations for either preventative maintenance or troubleshooting.

We therefore describes an approach towards asset management in a railroad transportation infrastructure owner and operator. Our approach focuses on the operational processes of asset maintenance and delivers immediate results by rightsizing and optimizing the allocation of the workforce across shifts and maintenance base locations. It is particularly helpful in a resource-constrained environment where managers believe a shortage of manpower and a backlog of work already exists. It can help everyone involved make the link in practical terms between personnel needs and an accounting for required maintenance tasks and assets.

This is zero-based headcount assignment approach makes no specific assumptions about existing personnel allocation. Under this modelling approach, each employee-hour allocated to a specific shift at a specific maintenance base must be accounted for, either assigned to overhead activities, unavoidable delays, or productive work. The model does not treat assignment of existing personnel as an input; it is based on documented maintenance work currently carried out by signal personnel, regardless of how they are covered by assignments.

We believe the classic approach towards asset management introduces a disconnect between theory and practice, leaving field forces confused as to why they are tasked to collect all of this data. Essentially by reversing the normal framework, we achieve much better buy-in and can show immediate results in terms of improving workforce management, followed by multiple stages of data refinement which ultimately result in a list of maintenance tasks that must be performed periodically, together with components to which they are applicable, which then rolls up to assets.

## Abridged Literature Review

Other researchers have recently developed optimization models to address signal maintenance work schedule assignments on an individual basis [5], however, it is not the goal of the approach described in this paper to determine individual assignments and their optimal sequence. Our framework is designed to elicit the appropriate overall staffing level in aggregate based on structured and validated inputs from maintenance managers.

A body of research also exists to explicitly address the trade-off between maintenance actions and probability of failure (e.g. [6]). That approach uses field-collected data to generate maintenance actions based on some optimization criteria (e.g. risk minimization). However, it assumes a constant available level of maintenance resources without analytically determining what is the appropriate level of maintenance input. It should be possible to utilize that approach to solve our current problem (e.g. by solving for maintenance resource requirements subject to a maximum risk constraint) given availability of extensive asset condition data, but that is not the approach we chose for this work. Our approach assumes a real-world condition where asset data is difficult to obtain, often unstructured, and discovers required maintenance actions utilizing maintenance managers' engineering judgment (including periodic maintenance, which usually isn't driven by asset condition but by regulation, and repair maintenance burden, based on historical trouble-call probabilities), and uses this elicited data to analytically determine the appropriate level of maintenance inputs and personnel assignments.

## Methods

We typically begin our rightsizing/resource review exercise by arranging a "day in the life of" visit with the maintenance line managers in the appropriate areas. At this point we will have already agreed with senior management as to scope of the effort, and they provide necessary introductions to the appropriate managers at their maintenance base (headquarters) facilities. When discussing these studies outside the corporate setting, we typically refer to them as "manning studies"; use of appropriate language is the first step towards gaining trust, and can suggest that we are there to listen rather than to tell them how to do their jobs.

### *Maintenance Employee Basics*

In a railroad environment, unionized (termed "agreement") maintenance employee positions are typically organized along four attributes: craft, location, gang, and shift. Craft refers to the defined set of tasks that an employee has knowledge, skills, abilities, and appropriate equipment and tools (personal, or issued by the railroad) to perform. Location is where the employees typically report for work at the beginning of each shift, usually governed by their collective bargaining agreements (CBAs). The CBAs may allow variations in reporting sites, but administratively they are assigned a location for purposes of collecting their weekly paycheck. Gang is the basic unit of workforce, typically comprising of several employees led by a team leader (Foreman), all of whom travel together in a group to and from the job site. Shared equipment may be assigned to a gang, making them responsible for looking after the equipment.

Maintenance forces are typically dispatched (both for preventative maintenance and trouble repair) by a district supervisor or manager, who may be responsible for multiple (headcount) positions assigned within a geographical area. It is at this level that we perform most of the work; the rightsizing study ensures that each supervisor has the appropriate personnel assigned to their district for the workload, and ensures that they are assigned the correct positions, which are centrally administered due to the nature of collective bargaining process. Typically the district supervisor has little control over which specific employees are assigned to which positions, and they can only change the craft, location, gang, and shift associated with their positions by administratively requesting it, in a process known as "abolishing" or "establishing" positions. Each round of position revisions may trigger a round of general workforce movement as eligible employees "bid" for open positions, or senior employees "bump" other employees out of their current positions. This "bid and bump" system was created in the early 20th century essentially to fairly distribute commutation burden and overtime opportunities to craft employees, all of whom are qualified to the same level and theoretically could perform any job requiring their craft.

### *Base Workload Data Collection*

Maintenance managers organize their workforce in different ways. Most have a "daily sheet" showing each employee's assigned activity for the day, including any employees out sick or on vacation. Some will use this to drive a "morning meeting" when Foremen in each gang are told of their assignments. This sheet, when collected over multiple days and discussed each day with the supervisor, provides a great deal of information about the group's current workload and personnel needs. A good time to discuss this with the supervisor is about half-way through the shift, when all gangs are working productively at their job sites and all daily workflow issues have been resolved, but prior to the busy period near end of the shift when overrunning work and overtime issues have to be sorted.

In the case of the Signal workforce, a large fraction of their workload is dedicated to Federally-mandated (FRA) signal apparatus inspections and tests. These tests have well-defined procedures (a Federal requirement), required frequencies defined in Code of Federal Regulations (49 CFR Part 234, 236), and must occur at defined field locations. To assess this work, we received copies of the Signal dept. procedures governing the "making of tests" (known as C&S 2, C&S 27 on various railroads), a work-assignment schedule used internally by the Signal dept. showing which gangs are assigned to which tests at which field locations, and a set of "block plans" which are logical descriptions of the signal system showing the location of all signal apparatus. We also received blank copies of forms used by the Signal workforce when making tests, which were helpful in interpreting the standard test procedures. At the same time, we received overview-level training on the Signal dept. procedures. In addition, we gathered a few sample filled-in forms from signal supervisors, which provided a means to verify our assessment of the time taken to perform each task.

Figure 1 shows a fictional sample of what a Signal dept. work assignment schedule looks like, for several locations hosting different types of signal apparatus. As can be seen, the required test at each location changes depending on what apparatus is installed there. Different tests at the same field location may also be assigned to different crafts, maintenance base locations, gangs, or shifts. Reasons for these vary, but typically it is due to different skills and equipment issued to different crafts. Local Signal Maintainers are able to perform most tests mechanical and electrical, but specialized relay testing is performed by Test Maintainers, who work out of a central relay shop within the district. Testing of electronic equipment, such as overlay track circuits, are performed by Electronic Technicians who have specialist equipment like oscilloscopes and advanced training with equipment vendors on the specifics of the electronic circuitry, like ability to change out circuit boards or replace electronically programmable read-only memory (EPROM) integrated circuit chips.

LOCATION : North Line Milepost 111.0 Control Point A

| Test | Freq | Test by | Test | Freq | Test by |
|---|---|---|---|---|---|
| 15 Switch Indication | 2 Yr | Gang #1 | 10 Switch Obst/Fouling Wires | 1 Mo | Gang #1 |
| 44 Track Circuits | 1 Yr | Gang #1 | 11 Point Detectors | 3 Mo | Gang #1 |
| 52A Timing Devices | 1 Yr | Gang #1 | 18 Fouling Circuits | 3 Mo | Gang #1 |
| 52B Time Locking (Intrlkng) | 2 Yr | Gang #1 | 30A Ground Readings | 1 Mo | Gang #1 |
| 77 Surge Protection Devices | 1 Yr | Gang #1 | 41 Insulated Joints | 3 Mo | Gang #1 |
| 78A Resistance / Made Grounds | 4 Yr | Gang #1 | | | |

LOCATION : North Line Milepost 139.8 Control Point B

| Test | Freq | Test by | Test | Freq | Test by |
|---|---|---|---|---|---|
| 15 Switch Indication | 2 Yr | Gang #3 | 10 Switch Obst/Fouling Wires | 1 Mo | Gang #3 |
| 44 Track Circuits | 1 Yr | Gang #3 | 11 Point Detectors | 3 Mo | Gang #3 |
| 52A Timing Devices | 1 Yr | Gang #3 | 18 Fouling Circuits | 3 Mo | Gang #3 |
| 52B Time Locking (Intrlkng) | 2 Yr | Gang #3 | 30A Ground Readings | 1 Mo | Gang #3 |
| 77 Surge Protection Devices | 1 Yr | Gang #3 | 41 Insulated Joints | 3 Mo | Gang #3 |
| 78A Resistance / Made Grounds | 4 Yr | Gang #3 | | | |
| BATT "C-CAN" Battery Change | 4 Yr | Gang #3 | | | |

LOCATION : North Line Milepost 141.3 Grade Crossing

| Test | Freq | Test by | Test | Freq | Test by |
|---|---|---|---|---|---|
| 26C Highway Crossing | 1 Yr | Gang #3 | 26A Grade Crossings | 1 Mo | Gang #3 |
| 78A Resistance / Made Grounds | 4 Yr | Gang #3 | 26B Grade Crossings | 3 Mo | Gang #3 |
| | | | 26H Hold Clear Check | 3 Mo | Gang #3 |

LOCATION : East Line Milepost 229.9

| Test | Freq | Test by | Test | Freq | Test by |
|---|---|---|---|---|---|
| 31D All Other Relays | 4 Yr | Test Gang | 65A Moveable Bridges | 3 Mo | Gang #4 |
| 32 Insulation/Resistance | 10 Yr | Test Gang | | | |
| 65B Moveable Bridges | 1 Yr | Gang #4 | | | |
| 77 Surge Protection Devices | 1 Yr | Gang #4 | | | |

LOCATION : North West Line Milepost 437.3 Master Location 4373

| Test | Freq | Test by |
|---|---|---|
| 31D All Other Relays | 4 Yr | Test Gang |
| 32 Insulation/Resistance | 10 Yr | Test Gang |
| SCT Special Cable Test | 1 Yr | Test Gang |

LOCATION : North West Line Milepost 469.3 Interlocking CP H

| Test | Freq | Test by |
|---|---|---|
| 31C1 AC Vane / DC Pole | 2 Yr | Test Gang |
| 31C2 Semaphore Mech Test | 2 Yr | Test Gang |
| 31D All Other Relays | 4 Yr | Test Gang |
| 32 Insulation/Resistance | 10 Yr | Test Gang |
| 50 Approach Locking | 2 Yr | Test Gang |
| 51 Signal Indication Locking | 2 Yr | Test Gang |
| 53 Route Locking | 2 Yr | Test Gang |

LOCATION : North West Line Milepost 439.8 Hot Box Detector

| Test | Freq | Test by | Test | Freq | Test by |
|---|---|---|---|---|---|
| 76 Overlays | 1 Yr | Elec. Tech. | 70A Hot Box Detector | 1 Mo | Elec. Tech. |
| 77 Surge Protection Devices | 1 Yr | Elec. Tech. | 70B Hot Box Detector | 3 Mo | Elec. Tech. |
| 78B Resistance / Made Grounds | 10 Yr | Elec. Tech. | 71 Dragging Eqp / 3rd Rail | 1 Mo | Elec. Tech. |

LOCATION : Central Line Milepost 6.9 Control Point Q

| Test | Freq | Test by | Test | Freq | Test by |
|---|---|---|---|---|---|
| 15 Switch Indication | 2 Yr | Gang #10 | 10 Switch Obst/Fouling Wires | 1 Mo | Gang #12 |
| 31C1 AC Vane / DC Pole | 2 Yr | Gang #10 | 11 Point Detectors | 3 Mo | Gang #12 |
| 31D All Other Relays | 4 Yr | Gang #10 | 18 Fouling Circuits | 3 Mo | Gang #12 |
| 32 Insulation/Resistance | 10 Yr | Gang #10 | 30A Ground Readings | 1 Mo | Gang #12 |
| 44A Cab Cut-in Circuit | 1 Yr | Gang #12 | 41 Insulated Joints | 3 Mo | Gang #12 |
| 500 Alternative Locking | 4 Yr | Elec. Tech. | CI Clutch Inspection | 6 Mo | Gang #12 |
| 52B Time Locking (Intrlkng) | 2 Yr | Gang #10 | | | |
| 53 Route Locking | 2 Yr | Gang #10 | | | |
| 77 Surge Protection Devices | 1 Yr | Gang #12 | | | |
| 78A Resistance / Made Grounds | 4 Yr | Gang #10 | | | |
| BC Bulb Change | 5 Yr | Gang #12 | | | |

**Figure 1.** Sample (Fictitious) Signal Dept. Work Schedules.

### *Base Workload Computation*

To assess the base workload, we held meetings with maintenance managers and supervisors to agree upon standardized timings and minimum crew to perform each test defined in the book. Minimum crew takes into account of constraints relating to minimum gang size, which might be contractually determined, and a matter of practical track safety (e.g. to provide lookout protection when working under traffic). Normally, we would separately assess the travel time to/from the site, coordination time (required to e.g. obtain necessary track outages to perform work), and necessary time to perform actual maintenance work. However, we chose a more aggregate approach for this assessment for several reasons:

1. Many of the specific tests themselves are very quick to perform by skilled personnel if no exceptions are taken, e.g. a switch obstruction test takes less than ten minutes per switch if the maintainer only needs to crank the switch from normal to reverse and back, to verify everything is within tolerance and completes the necessary paperwork. However, two factors introduce unpredictability into this process: (a) active traffic moving on the railroad, which needs to use the switches, determine when maintainers can gain access to each switch to perform the test, which in turn is determined by dispatchers whose primary responsibility is to keep traffic moving; (b) if a test reveals that components are out of tolerance, the issues must be repaired immediately or the switch must be taken out of service, but because of the intensive service on this railway, maintainers are expected to immediately make appropriate adjustments to return the component into a serviceable state, and this adjustment can take much longer, e.g. up to one hour.

2. The Signal dept. has many maintenance base locations throughout the entire territory, often situated in former signal tower locations between two to ten miles apart. In effect, signal maintainers serve as "on-site presence" for railroad infrastructure, and are often the first responders for situations not necessarily Signal dept. responsibilities, such as trespasser reports, or fire and smoke conditions. Although access to some of those locations are by no means easy (e.g. walking across complex interlocking plants under active traffic, walking in tunnels or along rights-of-way, or driving up to ten miles out of the way to cross the tracks on an overbridge or grade crossing, only to drive back to the original location on the other side), unlike other departments, the travel time to and from signal apparatus locations, on the order of ten minutes rather than an hour, tends to be a smaller fraction of the total time required for work.
3. The amount of work required for a given test at different locations vary, depending on local factors and day-to-day variations. e.g., for switch testing, a two-person crew could typically get through a small interlocking (up to six switches) in a half-day's worth of work, but even if that half-day's assignment consists of an interlocking with only one switch machine, there is still significant mobilization and demobilization time at the site, hard-to-predict amount of time spent taking switches out of service, plus some allowance in case that one switch requires adjustment to pass the test. Maintenance managers typically do not assign more than one location per half-day to ensure that sufficient time is available to properly carry out the work. Experience [3] has shown that when too many locations are assigned in too short a time period, incentives are introduced that can result in falsification of test reports, which has safety implications.

To implement this aggregate approach, we built a matrix of standard, all-inclusive work times for each type of test at each type of location. A small sample of this matrix is shown in Figure 2. Some tests are always completed in tandem with other tests, and are shown as "add-ons" where no additional travel/mobilization time is provided.

| Signal Apparatus Test | Location Type | | | | |
|---|---|---|---|---|---|
| | 1 | 2 | 3 | 4 | 5 |
| 10 Switch Obst/Fouling Wires & | | | | | |
| 11 Point Detectors | 0.5 | — | 0.5 | 0.5 | 1 |
| 16 Electric Lock on Hand Operated Swit | 0.33 | — | 0.33 | — | — |
| 18 Fouling Circuits | 0.5 | — | 0.5 | 0.5 | 1 |
| 26A/26B Grade Crossings | 0.5 | 0.5 | — | 0.5 | — |
| 31A Signal Mech Inspection | 0.5 | — | — | 0.5 | 1 |
| 31C1 AC Vane / DC Pole | 0.5 | 0.5 | 0.5 | 0.5 | 1 |
| 31C2 Semaphore Mech Test | 0.25 | — | — | 0.5 | 1 |
| 31D All Other Relays | 0.5 | 1 | 0.5 | 1 | 3 |
| 32 Insulation/Resistance | 0.5 | 0.5 | 0.5 | 1 | 1.5 |
| 41 Insulated Joints | 0.25 | 0.5 | 0.5 | 0.5 | 1 |
| 44 Track Circuits | — | — | 0.5 | 0.5 | 1 |
| 50 Approach Locking | — | — | — | 0.5 | 1 |
| 500 Alternative Locking | — | — | — | 0.5 | 1 |
| 52A Timing Devices | 0.5 | 0.5 | 0.5 | 0.5 | 1 |
| 52B Time Locking (Interlocking) | — | — | 0.5 | 0.5 | 1 |
| 53 Route Locking | — | — | — | 0.5 | 1.5 |
| 54 Traffic Locking | — | — | — | 0.5 | 1 |
| 55 Interlocking Machine | 0.5 | 0.5 | 0.5 | 0.5 | 0.5 |
| 65A Moveable Bridges | 0.5 | 0.5 | 0.5 | 0.5 | 0.5 |
| 65B Moveable Bridges | 0.5 | 0.5 | 0.5 | 0.5 | 0.5 |
| 76 Overlays | 0.5 | 0.5 | 0.5 | 0.5 | 1 |
| 77 Surge Protection Devices | 0.125 | 0.125 | 0.125 | 0.125 | 0.125 |

**Location Types:**
1. Code Change Point/Cut Section (Slave Location)/Master Location
2. Grade Crossing
3. Hand Operated Sw itch
4. Small Interlocking (Five Sw itches or Few er)
5. Large Interlocking (Six Sw itches or More)

**Figure 2.** Sample Task Unit Time Matrix (in Days) for Federally-Mandated Signal Apparatus Tests.

The key requirement of this matrix is that it must be agreed upon with the maintenance managers. Typically, maintenance managers have various practical incentive to get this right based on their experience: having too much time could lead to over-assignment of manpower, which leads to issues because idle personnel with abundance of time will tend to create disciplinary issues in a field environment and increase accident risk exposure; having too little time will lead to recurring assignment of overtime, which creates headaches for maintenance managers who will have to explain their use. In any case, the maintenance managers are aware that these times are subject to verification with classic industrial engineering methodologies (e.g. time and motion studies) by our group; we verified the timings for some high-frequency, high-quantity tasks through field observation, and found them quite accurate on the whole.

Because the time allotment for inspection is dependent on the location type, it is necessary to classify each work location based on the apparatus installed therein. We did this by comparing the list of locations with the information shown on the block plans. For this process, it was necessary to learn at a rudimentary level how to read the block plans, and understand the relationship between different components, such as transmitters and receivers used for overlay track circuits at grade crossings.

Once the unit time matrix has been agreed upon, the required gang-hours computation is essentially done by multiplying out the work schedule with the unit time. We call this part of our calculations the "base workload model". At this point, if there are any reason to suspect that the work schedule is either incomplete or inaccurate (e.g. decommissioned assets may have not been properly removed from a work schedule), a cross-check can be done with record drawings of the assets. We checked the work schedule against the block plans to ensure all locations were included, and against the track charts to ensure the correct number of switch machines and other components were shown. A further level of validation is possible by checking that each location receives all of the tests required based on the apparatus present in that location, although for the purposes of this study we assumed that the work schedules that are currently utilized operationally by the Signal dept. is correct in that regard.

***Trouble Ticket Workload***

Aside from Federally mandated tests and inspections, the Signal dept. has a variety of other workloads that are more difficult to quantify, but are nonetheless an important part of Signal maintenance work and must be properly manned. These include:

- Trouble Ticket (Repair Maintenance)
- Non-Federally Mandated Preventative Maintenance
- Technician-Initiated Corrective Maintenance
- Major Repairs and Non-Capital Renewal
- Coverage Duties
- Administrative and Overhead

To assess the trouble ticket workload, we gathered data from the Signal Control Desk, which is a part of the operations control centre responsible for logging and documenting signal repairs in real-time, and assisting Dispatchers in arranging for signal repairs outside normal office hours. The statistics provided included extensive data on signal failures by location, subsystem, and cause, for the entirety of 2016 over the whole territory. An anonymized view of this data is shown in Figure 3. Based on this data, we were able to assess trouble ticket volume by maintenance base location by associating asset location with their assigned maintenance bases. This was time- and day-of-week dependent as some maintenance base locations are unstaffed during certain shifts or at the weekend. The larger districts in the off-hours also affected travel time.

| | Fault Type | | | | | | | | | | | | | | | |
|---|---|---|---|---|---|---|---|---|---|---|---|---|---|---|---|---|
| **Location** | **1** | **2** | **3** | **4** | **5** | **6** | **7** | **8** | **9** | **10** | **11** | **12** | **13** | **14** | **15** | **Total** |
| **Interlocking A** | 23 | 7 | 0 | 44 | 2 | 0 | 0 | 0 | 0 | 0 | 0 | 1 | 1 | 0 | 4 | **82** |
| **Line Segment A-B** | 1 | 0 | 3 | 8 | 0 | 0 | 0 | 0 | 0 | 0 | 0 | 0 | 2 | 0 | 0 | **14** |
| **Interlocking AA** | 11 | 6 | 0 | 12 | 0 | 0 | 0 | 0 | 0 | 0 | 0 | 0 | 5 | 0 | 7 | **41** |
| **Interlocking B** | 12 | 8 | 0 | 22 | 0 | 0 | 0 | 0 | 0 | 0 | 0 | 2 | 3 | 0 | 6 | **53** |
| **Line Segment B-C** | 1 | 0 | 2 | 5 | 0 | 0 | 0 | 0 | 0 | 0 | 0 | 0 | 17 | 0 | 2 | **27** |
| **Interlocking C** | 5 | 3 | 0 | 10 | 0 | 0 | 0 | 0 | 43 | 0 | 0 | 0 | 0 | 0 | 3 | **64** |
| **Line Segment C-D** | 0 | 0 | 0 | 8 | 0 | 0 | 0 | 0 | 0 | 0 | 0 | 0 | 0 | 0 | 1 | **9** |
| **Interlocking D** | 5 | 0 | 0 | 13 | 0 | 0 | 0 | 0 | 0 | 0 | 0 | 0 | 2 | 0 | 5 | **25** |
| **Line Segment D-AA** | 0 | 0 | 0 | 4 | 0 | 0 | 0 | 0 | 0 | 0 | 0 | 1 | 1 | 0 | 1 | **7** |
| **Line Segment D-E** | 0 | 0 | 0 | 3 | 0 | 0 | 0 | 0 | 0 | 0 | 0 | 1 | 1 | 0 | 1 | **6** |
| **Interlocking E** | 3 | 1 | 1 | 9 | 0 | 0 | 0 | 0 | 0 | 0 | 0 | 0 | 1 | 1 | 5 | **21** |
| **Line Segment E-F** | 0 | 0 | 0 | 1 | 0 | 0 | 0 | 0 | 0 | 0 | 0 | 0 | 1 | 0 | 2 | **4** |
| **Terminal A** | 11 | 5 | 0 | 18 | 0 | 0 | 0 | 0 | 0 | 0 | 0 | 0 | 12 | 4 | 6 | **56** |
| **Terminal B** | 45 | 22 | 0 | 52 | 0 | 0 | 0 | 0 | 0 | 0 | 0 | 0 | 1 | 0 | 6 | **126** |
| **Central Office** | 0 | 0 | 0 | 0 | 0 | 0 | 0 | 2 | 0 | 0 | 11 | 10 | 1 | 0 | 7 | **31** |

| | | | | | |
|---|---|---|---|---|---|
| Fault Types | 1 Switch | 2 Signal | 3 Cab Signal | 4 Track Circuit | 5 Track Block |
| | 6 Switch Block | 7 Fleeting | 8 Traffic | 9 Bridge Span | 10 Grade Crossing |
| | 11 CTC | 12 Communications | 13 Detector | 14 Signal Power | 15 Other |

**Figure 3.** Sample Failure Occurrence Data by Location and Type.

However, no specific information was provided on the personnel required to resolve each issue. To understand the employees required for troubleshooting and repair work, we retrieved one month's worth of written reports consisting of 113 discrete events covering most categories, and from this we were able to quantitatively assess how many employees in which crafts were required to close each type of trouble ticket. For those categories for which we did not have historical data, we relied upon expert opinions of maintenance managers. Multiplying the trouble ticket volume by unit time yields gang-hours typically committed to resolving trouble tickets. We call this part of our calculations the "trouble ticket model".

Based on the timestamps of when the trouble tickets were opened and closed, we were also able to determine various statistical information:

- **percentage of trouble tickets worked by each shift** (for trouble ticket remaining open across multiple shifts, each impacted shift was included). This was important for allocating trouble ticket workload across shifts.
- **fraction of total hours by shift, during the rush-periods, and off-peak periods, when a trouble ticket remained open.** This is needed to determine to what extent the trouble ticket workload was committing employees which could otherwise be applied to preventative maintenance. The preventative work was only completed during the off-peak periods to avoid impacting revenue traffic during peak travel demand; thus, trouble ticket work utilizes otherwise unassignable personnel during rush-periods, but takes away forces from preventative maintenance during the off-peak. This computation was particularly complicated because the rush-hour work curfews varied by location, with peak-hour curfews at outer suburban locations occurring much earlier in the morning, and much later in the evening, compared to those locations in the city terminal district. Additionally, the overlap with rush and off-peak periods varied by shift, with all three weekday shifts impacted by one or both daily rush periods.

This statistical information is used in the process in converting failure counts to total maintenance burden, as detailed in Figure 4.

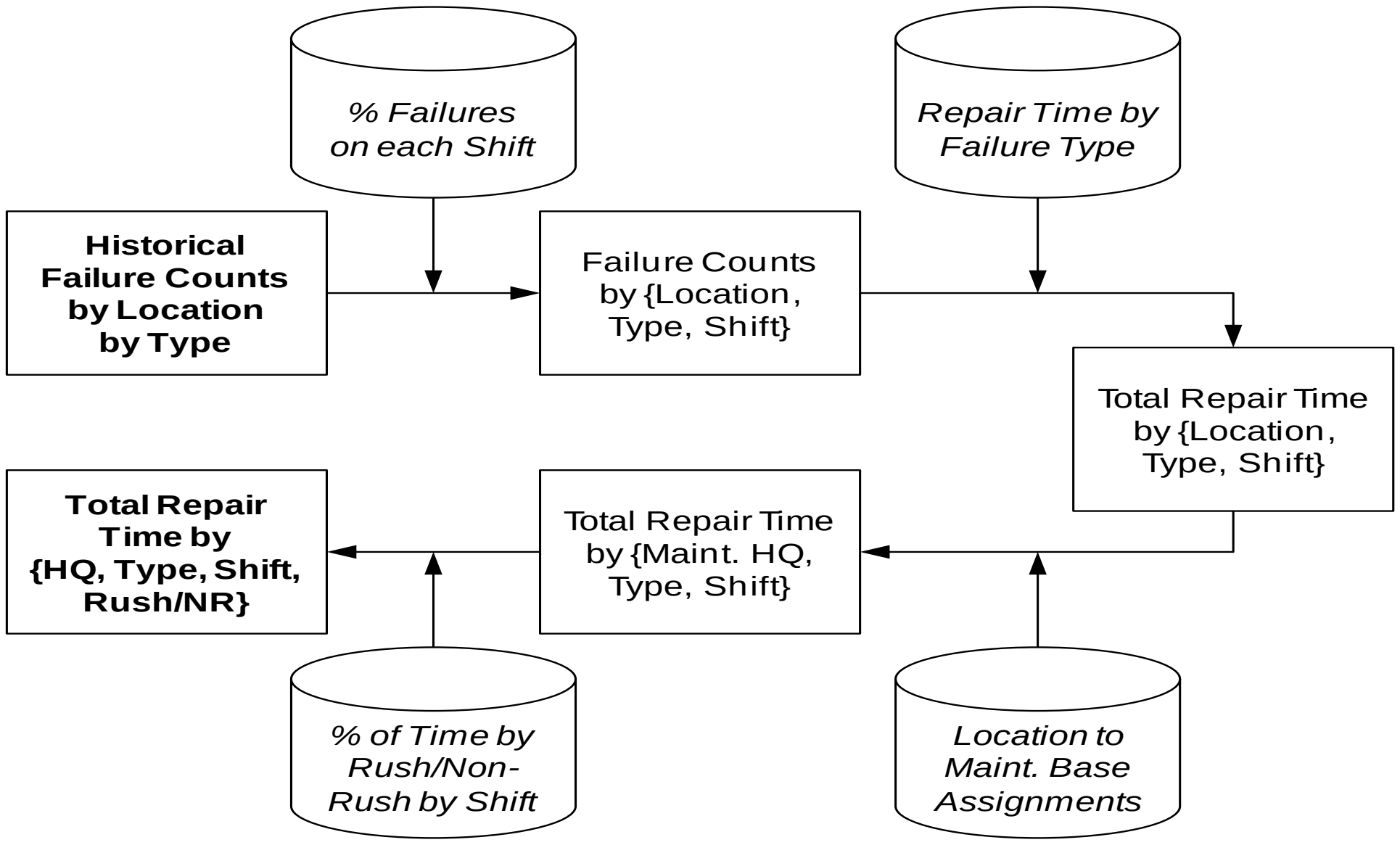

**Figure 4.** Block Diagram of the Trouble Ticket Model.

### *Non-Base, Non-Trouble Ticket (nBnTT) Workload*

The Non-Base, Non-Trouble Ticket workload captures all miscellaneous maintenance work that comprises part of infrastructure preventative maintenance but not explicitly required by the FRA. These workloads are necessary to maintain state of good repair, but are sometimes overlooked because they are not tracked by a regulatory compliance regime. We held a number of focus groups with maintenance managers to elicit as close as possible to a full inventory of all those tasks, the frequency with which they should be (or are typically) carried out, the unit-time required, and minimum crew for each task (Figure 5). We also reviewed a number of handwritten record books kept in each signal tower detailing work activities completed on each shift, to ensure that we properly canvassed all current maintenance work. We generated an inventory of signal assets by using work schedules previously utilized in computation of base workloads. Based on this information we computed the gang-hours requirements for this task scope.

| Task Description | For Each | Annual Occurrences | Work Time (Hours) | Crew Requirement |
|---|---|---|---|---|
| Location Inspection | Location | 4 | 1.5 | 2 |
| Paint Boxes | Location | 1 | 3 | 2 |
| Paint Switches/Signals | Interlocking | 4 | 8 | 2 |
| Oil Interlockings | Interlocking | 4 | 8 | 2 |
| Fall/Spring Switch Maint. | Interlocking | 2 | 8 | 2 |
| Snow Removal | Interlocking | 15 | 4 | 2 |
| Cold Weather Standby | Maint. Base | 30 | 7 | 2 |
| East Div. Snowmelter Tests | Interlocking | 26 | 1 | 2 |

| Task Description | For Each | Annual Occurrences | Work Time (Hours) | Crew Requirement |
|---|---|---|---|---|
| Weed Truck Support | Division | 1 | 8 | 1 |
| Graffiti Removal | Division | 52 | 6 | 2 |
| Maintain Solar Switches | Yard | 12 | 16 | 2 |
| Check Customer Reports | Maint. Base | 104 | 2 | 2 |
| Respond to Track Fires | Maint. Base | 9 | 4 | 2 |
| Power Outages | Maint. Base | 12 | 2 | 2 |
| Bridge Standby Duty | Bridge | 120 | 4 | 1 |

**Figure 5.** Sample Task Data for Non-Base, Non-Trouble Ticket Workload.

In certain cases, where Signal dept. workforce were required to support maintenance work "owned" by another area (e.g. Track, or Power, dept.), we determined based on the owning area's records the frequency, duration, and location of maintenance actions requiring Signal support. We also include "standby duty" as a task, which is when signal maintenance locations are staffed at a time when they normally wouldn't be, to reduce reliability risk and response time in the event of a failure, due to non-recurring but knowable-in-advance events, such as inclement weather and nice summer weekends (which dramatically increase the demand to operate movable bridges).

### *Assigning Workloads to Available Personnel and Work Windows*

The outputs from the three parts of workload assessment model discussed previously are in terms of gang-hours (inclusive of travel time, work time, and any process-induced idle time) at defined maintenance headquarter locations. However, the preventative work can occur on any shift (subject to timing constraints related to peak train traffic, and any work rule restrictions within collective bargaining agreements). It is therefore necessary to distribute the work to shifts whilst ensuring that each shift at each location has sufficient work window (time between rush periods) to complete the work, whilst minimizing the total amount of unassignable time. At the same time, the Signal dept. has definite preferences in terms of which location(s) and shift(s) to leave uncovered, if there is insufficient work to justify full coverage at all locations, 24-hours per day, and 7 days per week. For this reason, we call this part of the calculation the "coverage model". Figure 6 shows a high level block diagram of how this is accomplished.

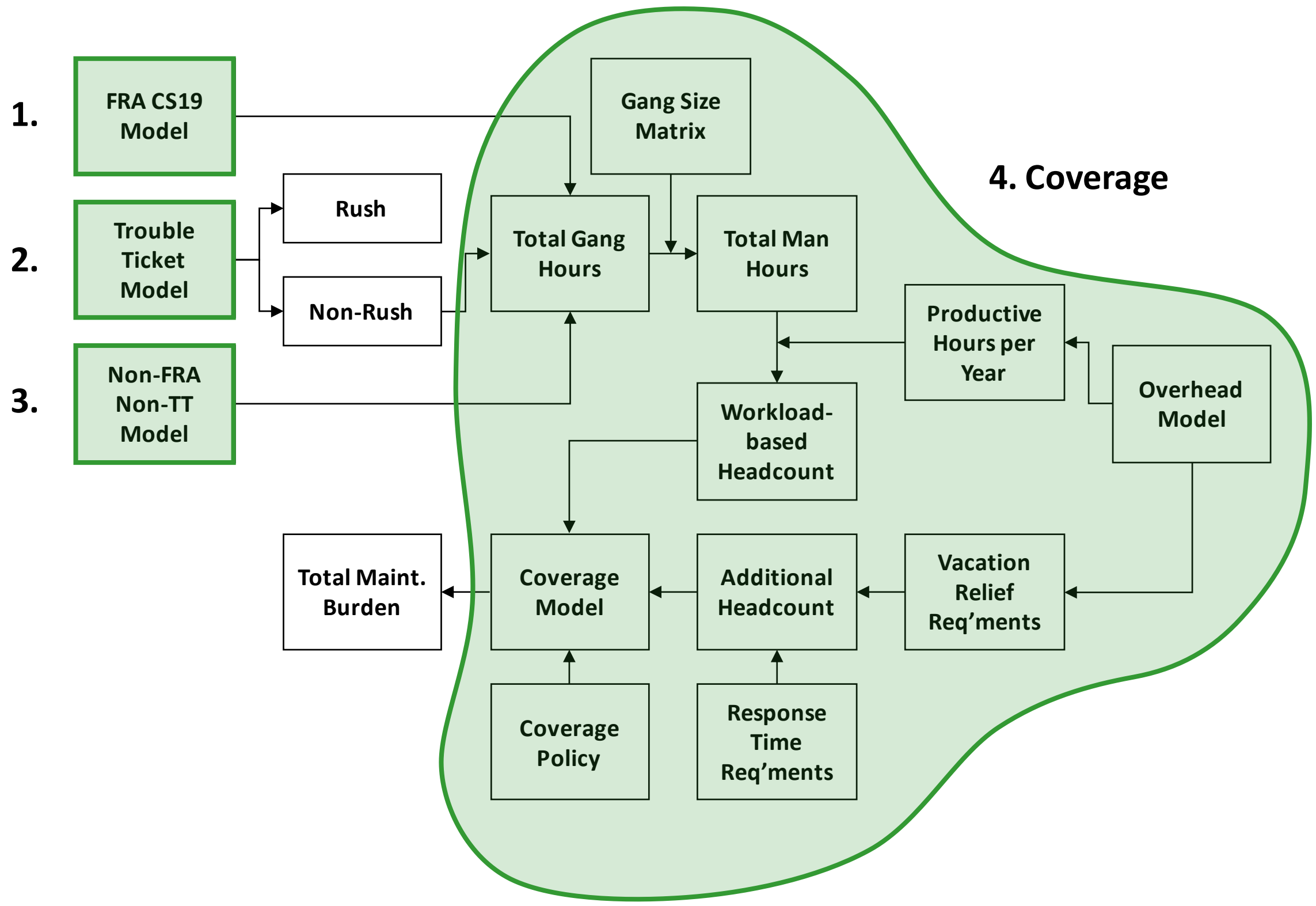

**Figure 6.** Overall Structure of Signal Dept. Workforce "Coverage" Model

It is assumed that all preventative maintenance work will be assigned outside of rush traffic period; from the trouble ticket model, only those workload attributed to off-peak period is included (during which they compete with preventative work for resourcing). We do not consider staffing needs during the rush periods, as only trouble ticket related repair maintenance work is authorized during that period (as not to disrupt revenue traffic), and more than adequate personnel is always available; employees are always available on standby duty at strategic locations as signal trouble during this period demands an immediate response to prevent traffic congestion from building up. Workload from all three areas are summed by maintenance base location. From this total gang-hour per location, based on the type of workload, gang-hours is multiplied by minimum gang size (by craft) for that type of work, which produces total required manhours by craft and by location (Figure 7).

| | Gang Hours | | | | | Man Hours | | | | Productive Hours/Yr | | | Full-Time Equivalents | | | | Alloted (Excl. Coverage) | | | |
|---|---|---|---|---|---|---|---|---|---|---|---|---|---|---|---|---|---|---|---|---|
| Tower | FRA Tests | Trouble Tickets | N/FRA Maint. | Elec. Tech. | Test Maint. | 1 | 2 | 3 | 4 | Non-Rush % | Maint. | All Other | 1 | 2 | 3 | 4 | 1 | 2 | 3 | 4 |
| DV | 1,490 | 207 | 952 | | | 4,822 | 348 | 310 | 0 | 56% | 1,025 | 1,034 | 4.70 | | | | 5 | | | |
| OW | 1,229 | 259 | 968 | | | 4,427 | 349 | 388 | 0 | 59% | 1,082 | 1,092 | 4.09 | | | | 4 | | | |
| HM | 3,129 | 422 | 1,332 | | | 9,101 | 657 | 633 | 0 | 58% | 1,054 | 1,063 | 8.64 | | | | 9 | | | |
| PS | | | | | 1,065 | 0 | 106 | 0 | 4,259 | 59% | 1,082 | 1,092 | | 1.95 | | 3.90 | | 2 | | 4 |
| PK | 2,936 | 439 | 1,571 | | | 9,107 | 670 | 658 | 0 | 64% | 1,168 | 1,178 | 7.80 | | | | 8 | | | |
| CA | 5,008 | 522 | 1,649 | | | 13,532 | 926 | 782 | 0 | 64% | 1,168 | 1,178 | 11.59 | 0.79 | | | 12 | 1 | | |
| CH | 2,818 | 688 | 1,132 | | | 8,712 | 739 | 1,033 | 0 | 59% | 1,082 | 1,092 | 8.05 | | | | 8 | | | |
| BR | 3,368 | 434 | 1,961 | | | 10,547 | 750 | 652 | 0 | 64% | 1,168 | 1,178 | 9.03 | 1.37 | | | 9 | 2 | | |
| ET | | | | 1,240 | | 0 | 124 | 3,721 | 0 | 58% | 1,054 | 1,063 | | | 7.69 | | | | 8 | |
| **North** | **19,979** | **2,971** | **9,565** | **1,240** | **1,065** | **60,248** | **4,670** | **8,178** | **4,259** | | | | **53.89** | **4.11** | **7.69** | **3.90** | **55** | **5** | **8** | **4** |
| SH | 995 | 374 | 954 | | | 4,169 | 382 | 562 | 0 | 56% | 1,025 | 1,034 | 4.07 | | | | 4 | | | |
| PK | 759 | 916 | 2,091 | | | 6,489 | 743 | 1,375 | 0 | 59% | 1,082 | 1,092 | 5.99 | | | | 6 | | | |
| ST | 2,064 | 676 | 1,405 | | | 7,587 | 685 | 1,014 | 0 | 59% | 1,082 | 1,092 | 7.01 | | | | 7 | | | |
| SD | 3,663 | | | 232 | | 7,325 | 389 | 696 | 0 | 59% | 1,082 | 1,092 | 6.77 | 2.01 | 3.34 | | 7 | 2 | 4 | |
| WK | 663 | 1,751 | 4,295 | | | 11,269 | 1,371 | 2,626 | 0 | 61% | 1,111 | 1,121 | 10.14 | | | | 10 | | | |
| CN | 1,216 | 458 | 1,218 | 128 | | 5,175 | 485 | 1,071 | 0 | 63% | 1,139 | 1,149 | 4.54 | 2.89 | 4.89 | | 5 | 3 | 5 | |
| DE | 1,240 | 404 | 1,755 | | | 5,919 | 501 | 605 | 0 | 66% | 1,196 | 1,207 | 4.95 | | | | 5 | | | |
| NH | 2,815 | 882 | 1,510 | | 962 | 9,658 | 970 | 1,323 | 3,848 | 66% | 1,196 | 1,207 | 8.07 | | | 3.52 | 8 | | | 4 |
| **East** | **13,413** | **5,461** | **13,228** | **360** | **962** | **57,590** | **5,527** | **9,272** | **3,848** | | | | **51.54** | **4.91** | **8.23** | **3.52** | **52** | **5** | **9** | **4** |
| TA | 1,055 | 429 | 706 | | | 4,026 | 390 | | 0 | 53% | 1,002 | 1,011 | 4.02 | 0.73 | | | 4 | 1 | | |
| TU | 169 | 479 | 918 | | | 2,673 | 348 | | 0 | 53% | 1,002 | 1,011 | 2.67 | | | | 3 | | | |
| NK | 421 | 284 | | | | 1,409 | 184 | | 0 | 56% | 1,025 | 1,034 | 1.37 | | | | 1 | | | |
| MO | 1,527 | 666 | 2,150 | 68 | 700 | 7,610 | 777 | 1,203 | 2,799 | 55% | 997 | 1,006 | 7.63 | 1.90 | 2.46 | 2.78 | 8 | 2 | 3 | 3 |
| WL | 1,190 | 465 | 1,048 | | | 4,880 | 456 | 697 | 0 | 58% | 1,054 | 1,063 | 4.63 | | | | 5 | | | |
| NW | 1,539 | 380 | 1,495 | | | 6,081 | 493 | 570 | 0 | 58% | 1,054 | 1,063 | 5.77 | | | | 6 | | | |
| **South** | **5,901** | **2,701** | **6,317** | **68** | **700** | **26,680** | **2,649** | **2,469** | **2,799** | | | | **26.09** | **2.63** | **2.46** | **2.78** | **27** | **3** | **3** | **3** |
| ***Total*** | ***39,292*** | ***11,134*** | ***29,110*** | ***1,668*** | ***2,727*** | ***144,517*** | ***12,847*** | ***19,919*** | ***10,907*** | | | | ***131.53*** | ***11.65*** | ***18.38*** | ***10.21*** | ***134*** | ***13*** | ***20*** | ***11*** |

Craft Legend
1. Maintainer or Assistant Inspector
2. Inspector
3. Electronic Technician
4. Test Maintainer

**Figure 7.** Converting Gang Hours to Allotted Positions for a Sample System.

To convert manhours into full-time equivalents (FTEs) that must be assigned at each location, it is necessary to assess the productive hours each FTE can provide. This is done in two stages. First the normal paid for no work (PFNW) items such as vacation, holidays, sick, personal days (defined by the collective bargaining agreement), training (per company policy), administrative overhead (e.g. discipline, random drug testing, idle time resulting from bid/bump procedures), and uncontrollable unproductive time based on historical occurrences (e.g. equipment malfunction, weather conditions) are subtracted from the number of weekdays in a year.

Secondly, the number of hours available to be assigned (per head) is multiplied the average percentage of each shift that is assignable for preventative work (i.e. outside of rush periods), which varies by location due to the different timing of peak traffic work curfew (Figure 8). The total number of manhours required by craft and location is divided by this estimate of productive hours per year. This produces, in essence, the minimum number of FTEs required at each location to ensure the work to be covered can be performed within the available work windows.

| Line | Loc. | Productive Hours | | | Night Shift (2300-0700) | | | | | | | | | | | | | | | | | | | | | | | | | | | | | | | |
|---|---|---|---|---|---|---|---|---|---|---|---|---|---|---|---|---|---|---|---|---|---|---|---|---|---|---|---|---|---|---|---|---|---|---|---|---|
| | | 1st Shift | 2nd Shift | Night Shift | 2300 | 2315 | 2330 | 2345 | 0000 | 0015 | 0030 | 0045 | 0100 | 0115 | 0130 | 0145 | 0200 | 0215 | 0230 | 0245 | 0300 | 0315 | 0330 | 0345 | 0400 | 0415 | 0430 | 0445 | 0500 | 0515 | 0530 | 0545 | 0600 | 0615 | 0630 | 0645 |
| South | TA | 4.75 | 3.75 | 7 | 0 | 0 | 0 | 0 | 0 | 0 | 0 | 0 | 0 | 0 | 0 | 0 | 0 | 0 | 0 | 0 | 0 | 0 | 0 | 0 | 0 | 0 | 0 | 0 | 0 | 0 | 0 | 0 | 1 | 1 | 1 | 1 |
| | NK | 5 | 4 | 7 | 0 | 0 | 0 | 0 | 0 | 0 | 0 | 0 | 0 | 0 | 0 | 0 | 0 | 0 | 0 | 0 | 0 | 0 | 0 | 0 | 0 | 0 | 0 | 0 | 0 | 0 | 0 | 0 | 1 | 1 | 1 | 1 |
| | MO | 5 | 3.75 | 6.75 | 0 | 0 | 0 | 0 | 0 | 0 | 0 | 0 | 0 | 0 | 0 | 0 | 0 | 0 | 0 | 0 | 0 | 0 | 0 | 0 | 0 | 0 | 0 | 0 | 0 | 0 | 0 | 1 | 1 | 1 | 1 | 1 |
| | WL | 5.25 | 4 | 6.75 | 0 | 0 | 0 | 0 | 0 | 0 | 0 | 0 | 0 | 0 | 0 | 0 | 0 | 0 | 0 | 0 | 0 | 0 | 0 | 0 | 0 | 0 | 0 | 0 | 0 | 0 | 0 | 1 | 1 | 1 | 1 | 1 |
| North | DV | 5.25 | 3.75 | 6.75 | 0 | 0 | 0 | 0 | 0 | 0 | 0 | 0 | 0 | 0 | 0 | 0 | 0 | 0 | 0 | 0 | 0 | 0 | 0 | 0 | 0 | 0 | 0 | 0 | 0 | 0 | 0 | 1 | 1 | 1 | 1 | 1 |
| | OW | 5.5 | 4 | 6.5 | 0 | 0 | 0 | 0 | 0 | 0 | 0 | 0 | 0 | 0 | 0 | 0 | 0 | 0 | 0 | 0 | 0 | 0 | 0 | 0 | 0 | 0 | 0 | 0 | 0 | 0 | 1 | 1 | 1 | 1 | 1 | 1 |
| | HM | 5.5 | 3.75 | 6.25 | 0 | 0 | 0 | 0 | 0 | 0 | 0 | 0 | 0 | 0 | 0 | 0 | 0 | 0 | 0 | 0 | 0 | 0 | 0 | 0 | 0 | 0 | 0 | 0 | 0 | 1 | 1 | 1 | 1 | 1 | 1 | 1 |
| | PS | 5.75 | 3.75 | 6 | 0 | 0 | 0 | 0 | 0 | 0 | 0 | 0 | 0 | 0 | 0 | 0 | 0 | 0 | 0 | 0 | 0 | 0 | 0 | 0 | 0 | 0 | 0 | 0 | 1 | 1 | 1 | 1 | 1 | 1 | 1 | 1 |
| | PK | 6.5 | 3.75 | 5.5 | 0 | 0 | 0 | 0 | 0 | 0 | 0 | 0 | 0 | 0 | 0 | 0 | 0 | 0 | 0 | 0 | 0 | 0 | 0 | 0 | 0 | 0 | 1 | 1 | 1 | 1 | 1 | 1 | 1 | 1 | 1 | 1 |
| | NW | 5.5 | 3.75 | 6.5 | 0 | 0 | 0 | 0 | 0 | 0 | 0 | 0 | 0 | 0 | 0 | 0 | 0 | 0 | 0 | 0 | 0 | 0 | 0 | 0 | 0 | 0 | 0 | 0 | 0 | 0 | 1 | 1 | 1 | 1 | 1 | 1 |
| | CH | 5.75 | 3.75 | 6 | 0 | 0 | 0 | 0 | 0 | 0 | 0 | 0 | 0 | 0 | 0 | 0 | 0 | 0 | 0 | 0 | 0 | 0 | 0 | 0 | 0 | 0 | 0 | 0 | 1 | 1 | 1 | 1 | 1 | 1 | 1 | 1 |
| | BR | 6.25 | 4 | 5.5 | 0 | 0 | 0 | 0 | 0 | 0 | 0 | 0 | 0 | 0 | 0 | 0 | 0 | 0 | 0 | 0 | 0 | 0 | 0 | 0 | 0 | 0 | 1 | 1 | 1 | 1 | 1 | 1 | 1 | 1 | 1 | 1 |
| East | SH | 5.25 | 3.75 | 6.5 | 0 | 0 | 0 | 0 | 0 | 0 | 0 | 0 | 0 | 0 | 0 | 0 | 0 | 0 | 0 | 0 | 0 | 0 | 0 | 0 | 0 | 0 | 0 | 0 | 0 | 0 | 1 | 1 | 1 | 1 | 1 | 1 |
| | PK | 5.5 | 4 | 6.5 | 0 | 0 | 0 | 0 | 0 | 0 | 0 | 0 | 0 | 0 | 0 | 0 | 0 | 0 | 0 | 0 | 0 | 0 | 0 | 0 | 0 | 0 | 0 | 0 | 0 | 0 | 1 | 1 | 1 | 1 | 1 | 1 |
| | ST | 5.75 | 3.75 | 6.25 | 0 | 0 | 0 | 0 | 0 | 0 | 0 | 0 | 0 | 0 | 0 | 0 | 0 | 0 | 0 | 0 | 0 | 0 | 0 | 0 | 0 | 0 | 0 | 0 | 0 | 1 | 1 | 1 | 1 | 1 | 1 | 1 |
| | SP | 6 | 3.75 | 6 | 0 | 0 | 0 | 0 | 0 | 0 | 0 | 0 | 0 | 0 | 0 | 0 | 0 | 0 | 0 | 0 | 0 | 0 | 0 | 0 | 0 | 0 | 0 | 0 | 1 | 1 | 1 | 1 | 1 | 1 | 1 | 1 |
| | WK | 6 | 3.75 | 6 | 0 | 0 | 0 | 0 | 0 | 0 | 0 | 0 | 0 | 0 | 0 | 0 | 0 | 0 | 0 | 0 | 0 | 0 | 0 | 0 | 0 | 0 | 0 | 0 | 1 | 1 | 1 | 1 | 1 | 1 | 1 | 1 |
| | CN | 6.25 | 3.75 | 5.5 | 0 | 0 | 0 | 0 | 0 | 0 | 0 | 0 | 0 | 0 | 0 | 0 | 0 | 0 | 0 | 0 | 0 | 0 | 0 | 0 | 0 | 0 | 1 | 1 | 1 | 1 | 1 | 1 | 1 | 1 | 1 | 1 |
| | DE | 6.5 | 4 | 5.25 | 0 | 0 | 0 | 0 | 0 | 0 | 0 | 0 | 0 | 0 | 0 | 0 | 0 | 0 | 0 | 0 | 0 | 0 | 0 | 0 | 0 | 1 | 1 | 1 | 1 | 1 | 1 | 1 | 1 | 1 | 1 | 1 |
| | NH | 6.75 | 3.75 | 5 | 0 | 0 | 0 | 0 | 0 | 0 | 0 | 0 | 0 | 0 | 0 | 0 | 0 | 0 | 0 | 0 | 0 | 0 | 0 | 0 | 1 | 1 | 1 | 1 | 1 | 1 | 1 | 1 | 1 | 1 | 1 | 1 |
| West | CA | 6.5 | 3.75 | 5.5 | 0 | 0 | 0 | 0 | 0 | 0 | 0 | 0 | 0 | 0 | 0 | 0 | 0 | 0 | 0 | 0 | 0 | 0 | 0 | 0 | 0 | 0 | 1 | 1 | 1 | 1 | 1 | 1 | 1 | 1 | 1 | 1 |

| Line | Loc. | First Shift (0700-1500) | | | | | | | | | | | | | | | | | | | | | | | | | | | | | | | |
|---|---|---|---|---|---|---|---|---|---|---|---|---|---|---|---|---|---|---|---|---|---|---|---|---|---|---|---|---|---|---|---|---|---|
| | | 0700 | 0715 | 0730 | 0745 | 0800 | 0815 | 0830 | 0845 | 0900 | 0915 | 0930 | 0945 | 1000 | 1015 | 1030 | 1045 | 1100 | 1115 | 1130 | 1145 | 1200 | 1215 | 1230 | 1245 | 1300 | 1315 | 1330 | 1345 | 1400 | 1415 | 1430 | 1445 |
| South | TA | 1 | 1 | 1 | 1 | 1 | 1 | 1 | 1 | 1 | 1 | 1 | 1 | 1 | 0 | 0 | 0 | 0 | 0 | 0 | 0 | 0 | 0 | 0 | 0 | 0 | 0 | 0 | 0 | 0 | 0 | 0 | 0 |
| | NK | 1 | 1 | 1 | 1 | 1 | 1 | 1 | 1 | 1 | 1 | 1 | 1 | 0 | 0 | 0 | 0 | 0 | 0 | 0 | 0 | 0 | 0 | 0 | 0 | 0 | 0 | 0 | 0 | 0 | 0 | 0 | 0 |
| | MO | 1 | 1 | 1 | 1 | 1 | 1 | 1 | 1 | 1 | 1 | 1 | 1 | 0 | 0 | 0 | 0 | 0 | 0 | 0 | 0 | 0 | 0 | 0 | 0 | 0 | 0 | 0 | 0 | 0 | 0 | 0 | 0 |
| | WL | 1 | 1 | 1 | 1 | 1 | 1 | 1 | 1 | 1 | 1 | 1 | 0 | 0 | 0 | 0 | 0 | 0 | 0 | 0 | 0 | 0 | 0 | 0 | 0 | 0 | 0 | 0 | 0 | 0 | 0 | 0 | 0 |
| North | DV | 1 | 1 | 1 | 1 | 1 | 1 | 1 | 1 | 1 | 1 | 1 | 0 | 0 | 0 | 0 | 0 | 0 | 0 | 0 | 0 | 0 | 0 | 0 | 0 | 0 | 0 | 0 | 0 | 0 | 0 | 0 | 0 |
| | OW | 1 | 1 | 1 | 1 | 1 | 1 | 1 | 1 | 1 | 1 | 0 | 0 | 0 | 0 | 0 | 0 | 0 | 0 | 0 | 0 | 0 | 0 | 0 | 0 | 0 | 0 | 0 | 0 | 0 | 0 | 0 | 0 |
| | HM | 1 | 1 | 1 | 1 | 1 | 1 | 1 | 1 | 1 | 1 | 0 | 0 | 0 | 0 | 0 | 0 | 0 | 0 | 0 | 0 | 0 | 0 | 0 | 0 | 0 | 0 | 0 | 0 | 0 | 0 | 0 | 0 |
| | PS | 1 | 1 | 1 | 1 | 1 | 1 | 1 | 1 | 1 | 0 | 0 | 0 | 0 | 0 | 0 | 0 | 0 | 0 | 0 | 0 | 0 | 0 | 0 | 0 | 0 | 0 | 0 | 0 | 0 | 0 | 0 | 0 |
| | PK | 1 | 1 | 1 | 1 | 1 | 1 | 0 | 0 | 0 | 0 | 0 | 0 | 0 | 0 | 0 | 0 | 0 | 0 | 0 | 0 | 0 | 0 | 0 | 0 | 0 | 0 | 0 | 0 | 0 | 0 | 0 | 0 |
| | NW | 1 | 1 | 1 | 1 | 1 | 1 | 1 | 1 | 1 | 1 | 0 | 0 | 0 | 0 | 0 | 0 | 0 | 0 | 0 | 0 | 0 | 0 | 0 | 0 | 0 | 0 | 0 | 0 | 0 | 0 | 0 | 0 |
| | CH | 1 | 1 | 1 | 1 | 1 | 1 | 1 | 1 | 1 | 0 | 0 | 0 | 0 | 0 | 0 | 0 | 0 | 0 | 0 | 0 | 0 | 0 | 0 | 0 | 0 | 0 | 0 | 0 | 0 | 0 | 0 | 0 |
| | BR | 1 | 1 | 1 | 1 | 1 | 1 | 1 | 0 | 0 | 0 | 0 | 0 | 0 | 0 | 0 | 0 | 0 | 0 | 0 | 0 | 0 | 0 | 0 | 0 | 0 | 0 | 0 | 0 | 0 | 0 | 0 | 0 |
| East | SH | 1 | 1 | 1 | 1 | 1 | 1 | 1 | 1 | 1 | 1 | 1 | 0 | 0 | 0 | 0 | 0 | 0 | 0 | 0 | 0 | 0 | 0 | 0 | 0 | 0 | 0 | 0 | 0 | 0 | 0 | 0 | 0 |
| | PK | 1 | 1 | 1 | 1 | 1 | 1 | 1 | 1 | 1 | 1 | 0 | 0 | 0 | 0 | 0 | 0 | 0 | 0 | 0 | 0 | 0 | 0 | 0 | 0 | 0 | 0 | 0 | 0 | 0 | 0 | 0 | 0 |
| | ST | 1 | 1 | 1 | 1 | 1 | 1 | 1 | 1 | 1 | 0 | 0 | 0 | 0 | 0 | 0 | 0 | 0 | 0 | 0 | 0 | 0 | 0 | 0 | 0 | 0 | 0 | 0 | 0 | 0 | 0 | 0 | 0 |
| | SP | 1 | 1 | 1 | 1 | 1 | 1 | 1 | 1 | 0 | 0 | 0 | 0 | 0 | 0 | 0 | 0 | 0 | 0 | 0 | 0 | 0 | 0 | 0 | 0 | 0 | 0 | 0 | 0 | 0 | 0 | 0 | 0 |
| | WK | 1 | 1 | 1 | 1 | 1 | 1 | 1 | 1 | 0 | 0 | 0 | 0 | 0 | 0 | 0 | 0 | 0 | 0 | 0 | 0 | 0 | 0 | 0 | 0 | 0 | 0 | 0 | 0 | 0 | 0 | 0 | 0 |
| | CN | 1 | 1 | 1 | 1 | 1 | 1 | 1 | 0 | 0 | 0 | 0 | 0 | 0 | 0 | 0 | 0 | 0 | 0 | 0 | 0 | 0 | 0 | 0 | 0 | 0 | 0 | 0 | 0 | 0 | 0 | 0 | 0 |
| | DE | 1 | 1 | 1 | 1 | 1 | 1 | 0 | 0 | 0 | 0 | 0 | 0 | 0 | 0 | 0 | 0 | 0 | 0 | 0 | 0 | 0 | 0 | 0 | 0 | 0 | 0 | 0 | 0 | 0 | 0 | 0 | 0 |
| | NH | 1 | 1 | 1 | 1 | 1 | 0 | 0 | 0 | 0 | 0 | 0 | 0 | 0 | 0 | 0 | 0 | 0 | 0 | 0 | 0 | 0 | 0 | 0 | 0 | 0 | 0 | 0 | 0 | 0 | 0 | 0 | 0 |
| West | CA | 1 | 1 | 1 | 1 | 1 | 1 | 0 | 0 | 0 | 0 | 0 | 0 | 0 | 0 | 0 | 0 | 0 | 0 | 0 | 0 | 0 | 0 | 0 | 0 | 0 | 0 | 0 | 0 | 0 | 0 | 0 | 0 |

| Line | Loc. | Second Shift (1500-2300) | | | | | | | | | | | | | | | | | | | | | | | | | | | | | | | |
|---|---|---|---|---|---|---|---|---|---|---|---|---|---|---|---|---|---|---|---|---|---|---|---|---|---|---|---|---|---|---|---|---|---|
| | | 1500 | 1515 | 1530 | 1545 | 1600 | 1615 | 1630 | 1645 | 1700 | 1715 | 1730 | 1745 | 1800 | 1815 | 1830 | 1845 | 1900 | 1915 | 1930 | 1945 | 2000 | 2015 | 2030 | 2045 | 2100 | 2115 | 2130 | 2145 | 2200 | 2215 | 2230 | 2245 |
| South | TA | 0 | 0 | 0 | 0 | 1 | 1 | 1 | 1 | 1 | 1 | 1 | 1 | 1 | 1 | 1 | 1 | 1 | 1 | 1 | 1 | 1 | 0 | 0 | 0 | 0 | 0 | 0 | 0 | 0 | 0 | 0 | 0 |
| | NK | 0 | 0 | 0 | 0 | 0 | 1 | 1 | 1 | 1 | 1 | 1 | 1 | 1 | 1 | 1 | 1 | 1 | 1 | 1 | 1 | 1 | 0 | 0 | 0 | 0 | 0 | 0 | 0 | 0 | 0 | 0 | 0 |
| | MO | 0 | 0 | 0 | 0 | 0 | 1 | 1 | 1 | 1 | 1 | 1 | 1 | 1 | 1 | 1 | 1 | 1 | 1 | 1 | 1 | 1 | 1 | 0 | 0 | 0 | 0 | 0 | 0 | 0 | 0 | 0 | 0 |
| | WL | 0 | 0 | 0 | 0 | 0 | 0 | 1 | 1 | 1 | 1 | 1 | 1 | 1 | 1 | 1 | 1 | 1 | 1 | 1 | 1 | 1 | 1 | 0 | 0 | 0 | 0 | 0 | 0 | 0 | 0 | 0 | 0 |
| North | DV | 0 | 0 | 0 | 0 | 0 | 0 | 1 | 1 | 1 | 1 | 1 | 1 | 1 | 1 | 1 | 1 | 1 | 1 | 1 | 1 | 1 | 1 | 1 | 0 | 0 | 0 | 0 | 0 | 0 | 0 | 0 | 0 |
| | OW | 0 | 0 | 0 | 0 | 0 | 0 | 0 | 0 | 1 | 1 | 1 | 1 | 1 | 1 | 1 | 1 | 1 | 1 | 1 | 1 | 1 | 1 | 1 | 1 | 0 | 0 | 0 | 0 | 0 | 0 | 0 | 0 |
| | HM | 0 | 0 | 0 | 0 | 0 | 0 | 0 | 0 | 1 | 1 | 1 | 1 | 1 | 1 | 1 | 1 | 1 | 1 | 1 | 1 | 1 | 1 | 1 | 1 | 1 | 0 | 0 | 0 | 0 | 0 | 0 | 0 |
| | PS | 0 | 0 | 0 | 0 | 0 | 0 | 0 | 0 | 0 | 1 | 1 | 1 | 1 | 1 | 1 | 1 | 1 | 1 | 1 | 1 | 1 | 1 | 1 | 1 | 1 | 1 | 0 | 0 | 0 | 0 | 0 | 0 |
| | PK | 0 | 0 | 0 | 0 | 0 | 0 | 0 | 0 | 0 | 0 | 0 | 1 | 1 | 1 | 1 | 1 | 1 | 1 | 1 | 1 | 1 | 1 | 1 | 1 | 1 | 1 | 1 | 1 | 0 | 0 | 0 | 0 |
| | NW | 0 | 0 | 0 | 0 | 0 | 0 | 1 | 1 | 1 | 1 | 1 | 1 | 1 | 1 | 1 | 1 | 1 | 1 | 1 | 1 | 1 | 1 | 1 | 0 | 0 | 0 | 0 | 0 | 0 | 0 | 0 | 0 |
| | CH | 0 | 0 | 0 | 0 | 0 | 0 | 0 | 1 | 1 | 1 | 1 | 1 | 1 | 1 | 1 | 1 | 1 | 1 | 1 | 1 | 1 | 1 | 1 | 1 | 0 | 0 | 0 | 0 | 0 | 0 | 0 | 0 |
| | BR | 0 | 0 | 0 | 0 | 0 | 0 | 0 | 0 | 0 | 0 | 0 | 1 | 1 | 1 | 1 | 1 | 1 | 1 | 1 | 1 | 1 | 1 | 1 | 1 | 1 | 1 | 1 | 0 | 0 | 0 | 0 | 0 |
| East | SH | 0 | 0 | 0 | 0 | 0 | 0 | 1 | 1 | 1 | 1 | 1 | 1 | 1 | 1 | 1 | 1 | 1 | 1 | 1 | 1 | 1 | 1 | 1 | 0 | 0 | 0 | 0 | 0 | 0 | 0 | 0 | 0 |
| | PK | 0 | 0 | 0 | 0 | 0 | 0 | 0 | 1 | 1 | 1 | 1 | 1 | 1 | 1 | 1 | 1 | 1 | 1 | 1 | 1 | 1 | 1 | 1 | 0 | 0 | 0 | 0 | 0 | 0 | 0 | 0 | 0 |
| | ST | 0 | 0 | 0 | 0 | 0 | 0 | 0 | 1 | 1 | 1 | 1 | 1 | 1 | 1 | 1 | 1 | 1 | 1 | 1 | 1 | 1 | 1 | 1 | 1 | 0 | 0 | 0 | 0 | 0 | 0 | 0 | 0 |
| | SP | 0 | 0 | 0 | 0 | 0 | 0 | 0 | 0 | 1 | 1 | 1 | 1 | 1 | 1 | 1 | 1 | 1 | 1 | 1 | 1 | 1 | 1 | 1 | 1 | 1 | 0 | 0 | 0 | 0 | 0 | 0 | 0 |
| | WK | 0 | 0 | 0 | 0 | 0 | 0 | 0 | 0 | 1 | 1 | 1 | 1 | 1 | 1 | 1 | 1 | 1 | 1 | 1 | 1 | 1 | 1 | 1 | 1 | 1 | 0 | 0 | 0 | 0 | 0 | 0 | 0 |
| | CN | 0 | 0 | 0 | 0 | 0 | 0 | 0 | 0 | 0 | 1 | 1 | 1 | 1 | 1 | 1 | 1 | 1 | 1 | 1 | 1 | 1 | 1 | 1 | 1 | 1 | 1 | 0 | 0 | 0 | 0 | 0 | 0 |
| | DE | 0 | 0 | 0 | 0 | 0 | 0 | 0 | 0 | 0 | 0 | 1 | 1 | 1 | 1 | 1 | 1 | 1 | 1 | 1 | 1 | 1 | 1 | 1 | 1 | 1 | 1 | 0 | 0 | 0 | 0 | 0 | 0 |
| | NH | 0 | 0 | 0 | 0 | 0 | 0 | 0 | 0 | 0 | 0 | 0 | 1 | 1 | 1 | 1 | 1 | 1 | 1 | 1 | 1 | 1 | 1 | 1 | 1 | 1 | 1 | 1 | 1 | 0 | 0 | 0 | 0 |
| West | CA | 0 | 0 | 0 | 0 | 0 | 0 | 0 | 0 | 0 | 0 | 1 | 1 | 1 | 1 | 1 | 1 | 1 | 1 | 1 | 1 | 1 | 1 | 1 | 1 | 1 | 1 | 1 | 0 | 0 | 0 | 0 | 0 |

**Figure 8.** Graphical Representation of Work Curfew Impacts on Maintenance Productivity

From the allotted FTEs at each location by craft, it is necessary to allocate them to shifts and meet the minimum gang requirements on each shift. This is in essence done manually based on a grid of assignment rules. Allotted headcount of 4 results in assignment of Monday through Friday, first and second shifts of two employees each; headcount of 8 results in first and second shift having 7-day coverage with three employees rotating rest days except for Wednesday when all three employees are present, whilst overnight shift is assigned Sunday through Thursday, to cover Monday through Friday morning rush period; allotted headcount of 7 results in 8 being assigned because there is no feasible shift assignment that would fulfill minimum gang size constraints. Figure 9 details example shift assignment rules for assistant inspectors and signal maintainers.

| Asst. Insp. Assignment (AM/PM/Night) | Maintainer Assignment (AM/PM/Night) | Total Headcount | Resulting Coverage |
|---|---|---|---|
| 1/0/0 | 1/2/0 | 4 | 2-person gangs M-F, AM and PM shifts |
| 1/0/0 | 1/2/2 | 6 | 2-person gangs M-F, 24-hours |
| 1/0/0 | 2/3/0 | 6 | 2-person gangs 7 days (3-person on Wed), AM and PM shifts (0700-1500 and 1500-2300) |
| 1/0/0 | 2/3/2 | 8 | 2-person gangs 7 days (3-person on Wed), AM and PM shifts; 2-person gangs Su-Th overnight |
| 1/0/0 | 2/3/3 | 9 | 2-person gangs 7 days (3-person on Wed), 24-hours |

**Figure 9.** Sample Assignment Rules to Provide Various Levels of Time and Day-of-Week Coverage

This work assignment process assume that no overtime is to be incurred, as overtime availability cannot be guaranteed and should not be relied to work fulfill base workload such as preventative maintenance. We plan on the basis of no overtime; the appropriate use of overtime is as an operational measure to recover from unforeseen contingencies such as a major failure, a work backlog, or lower than planned personnel or work window availability, e.g. due to retirement timing or conflicts with capital projects required by other depts.

### *Assigning Vacation Relief and Heavy Repair Gang Positions*

At this point, extra positions needed to be added to account for heavy repair gang (where specialized knowledge to operate heavy construction equipment is necessary, to deal with situations such as knocked-down grade crossings, tree removal, buried cable replacement, signal case installation and removals, etc.—one for each division, always assigned first shift), and for vacation relief (based on assigned positions on each shift and fraction of work days requiring coverage). When these positions are added to the assigned positions above, the result is total headcount required to ensure all assets receive all their preventative maintenance and sufficient employees are set aside to cover unexpected repairs and normal operating situations such as personnel absence. The results are produced at a sufficient level of granularity to inform position assignment decisions by craft, location, gang, and shift. As a biproduct, an asset inventory is generated, at a sufficient level of detail to assess maintenance requirements, and could feed normal asset management processes in terms of assessing replacement cycles, consumption of maintenance employee-hours, risk of asset failures, and the organization's ability to respond to them.

## RESULTS AND DISCUSSION

### *Understanding Employee-Hour Consumption by Constraint*

One of the outputs of this process is an accounting of time utilization of the average frontline worker in terms of overhead versus maintenance activity. We have built the model by assessing each task and "revenue hour" required to do them (where actively performing maintenance is considered a "revenue-generating" activity—from the perspective of an internal maintenance contractor), and added the "overhead hours" over several stages to account for different types of "non-revenue" hours. Thus, by inspection of this data and comparing the total hours required at each stage of the model, we arrive at an understanding of how different operational constraints contribute to the overall workforce requirements. This data could also be used to generate a breakdown to understand how much labour and supervisory hours each asset consumes, because each task is tied to an asset in the field. Over the longer term these statistics could be used for benchmarking or informing management strategic actions in driving these indicators one way or another.

Figure 10 shows allocation of time by division and by craft for a sample system. This system is heavily oriented towards commuter passenger traffic, as a result have restrictive work curfews during the morning and evening rush periods. As these statistics demonstrate, work curfews have a huge impact on the productivity of the workforce.

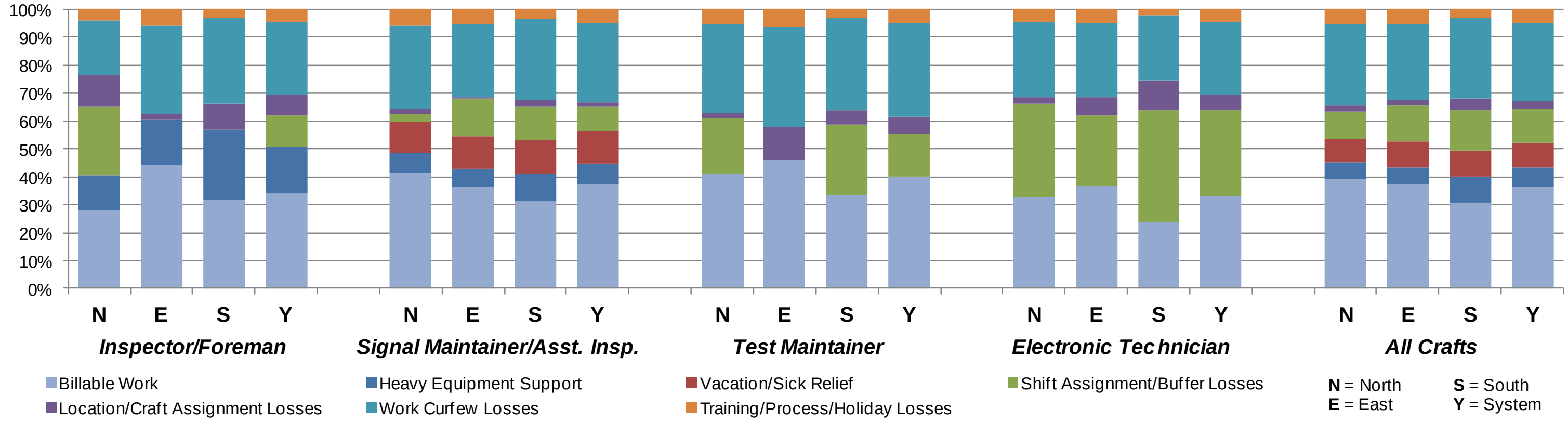

| Craft: | Inspector/Foreman | | | | Signal Maintainer/Asst. Insp. | | | | Test Maintainer | | | | Electronic Technician | | | | All Crafts | | | |
|---|---|---|---|---|---|---|---|---|---|---|---|---|---|---|---|---|---|---|---|---|
| **Time Allocation** Division: | North | East | South | Sys | North | East | South | Sys | North | East | South | Sys | North | East | South | Sys | North | East | South | Sys |
| Heavy Equipment Support | 12.5% | 16.7% | 25.0% | 16.7% | 7.1% | 6.6% | 9.8% | 7.5% | — | — | — | — | — | — | — | — | 6.3% | 6.1% | 9.3% | 6.9% |
| Vacation/Sick Relief | — | — | — | — | 11.4% | 11.8% | 12.2% | 11.8% | — | — | — | — | — | — | — | — | 8.4% | 9.2% | 9.3% | 8.9% |
| Shift Assignment/Buffer Losses | 25.0% | — | — | 11.1% | 2.9% | 13.2% | 12.2% | 9.1% | 20.0% | — | 25.0% | 15.4% | 33.3% | 25.0% | 40.0% | 31.0% | 9.5% | 13.3% | 14.8% | 12.1% |
| Location/Craft Assignment Losses | 11.2% | 1.5% | 9.2% | 7.5% | 1.6% | 0.6% | 2.2% | 1.3% | 2.0% | 11.9% | 5.4% | 6.1% | 2.6% | 6.4% | 10.9% | 5.6% | 2.5% | 1.8% | 3.8% | 2.5% |
| Work Curfew Losses | 19.6% | 31.7% | 31.0% | 26.2% | 29.8% | 26.3% | 29.2% | 28.2% | 31.7% | 35.8% | 32.8% | 33.3% | 27.0% | 26.6% | 23.2% | 26.2% | 28.7% | 27.0% | 29.0% | 28.1% |
| Training/Process/Holiday Losses | 3.8% | 6.0% | 3.1% | 4.3% | 6.0% | 5.3% | 3.3% | 5.1% | 5.5% | 6.2% | 3.2% | 5.0% | 4.4% | 5.0% | 2.3% | 4.3% | 5.6% | 5.3% | 3.2% | 4.9% |
| **Billable Work** | **28.0%** | **44.1%** | **31.7%** | **34.2%** | **41.2%** | **36.3%** | **31.2%** | **37.0%** | **40.8%** | **46.1%** | **33.5%** | **40.2%** | **32.7%** | **37.0%** | **23.7%** | **32.9%** | **39.0%** | **37.3%** | **30.7%** | **36.5%** |

**Figure 10.** Time Allocation Results for a Sample System

Another major consumer of employee-hours is shift assignment and buffer losses. Shift assignment is the process of ensuring that each location has at least two employees assigned at the times when the location is open, because Signal employees always work in teams of two (the two employee could be of different craft), due to task logistical requirements and for electrical and track safety reasons. Most locations have workloads that justify staffing more than first and second shift on weekdays (4 positions), but not many have workloads that require more than a crew of two on a rotating 24-7 schedule (9 positions). This means inevitably there exists much "fractional" workloads, which must be rounded up in the assignment process, unless locations are closed and the fractional work transferred to adjacent locations. As discussed earlier, consolidating locations will have significant response-time impacts during the commuter rush hours and for emergencies such as a track fires (particularly because commuter rush hours is also when the adjacent highway infrastructure needed to access signal infrastructure is extremely congested), so in practice this is very difficult to do.

There is also a common misconception that CBAs constrain to what extent maintenance activities can be optimized, and by doing away with them the operation would be instantaneously more efficient. In fact this data shows that the location/craft assignment process where most of the CBA-related constraints are applied accounts for only 2.5% of productivity losses, whereas rush period work curfew and shift assignments results in much more significant losses.

### *Employee-Hour Consumption by Task Category*

Another result is an understanding of how much time (at least in theory) is expended on each category of maintenance processes. It is not possible to know this without the model, because maintenance do not currently provide structured logging of their day-to-day maintenance activities (although it may be possible in theory to reconstruct this information from handwritten paper timesheets and record books). This informs us as how much we could hope to influence maintenance expenditures. The Federally mandated tests and inspections are periodic, and those are what economists would term "inelastic demand", in that better maintenance will never lead to reduction of this burden—thus the only way to reduce this burden is to reduce asset base. On the contrary, reduction of failure rates (by asset replacement, improved maintenance regime, etc.) might drive down the time spent on trouble tickets, but would increase Non-FRA maintenance burden.

Figure 11 shows time utilization by division and by maintenance base for a sample system. Some maintenance bases (e.g. PS, SD, WK, NK, TU—each code stands for an interlocking tower) specialize only in certain type of duties, leading to uneven distribution. However, on a division and systemwide basis, this sample system devotes just over half of its Signal maintenance hours to Federally mandated tests, about another quarter to non-FRA maintenance; only 25% are utilized in repair maintenance, indicating that the combined FRA and non-FRA maintenance regime is working well to proactively keep assets in a state of good repair. Expending a vast majority of maintenance resources on repairs would indicate a more reactive approach and possibly a maintenance backlog.

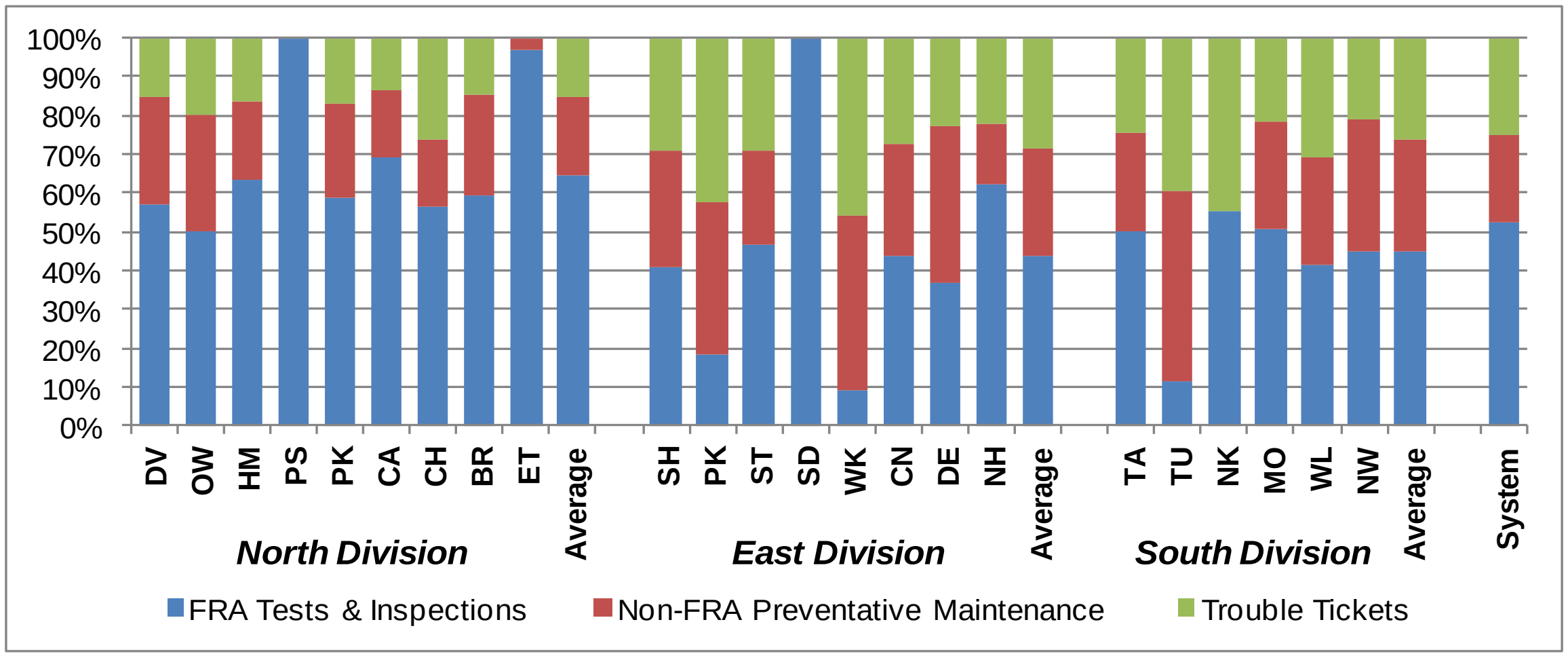

**Figure 11.** Time Utilization Results for a Sample System

### *Zero-based Personnel Assignment*

The most important result of this process, from the maintenance manager's point of view, is an accounting of how many employees of each craft should be assigned to each location on each shift, to ensure that there is sufficient employees to cover all necessary and anticipated workload. Figure 12 shows this position assignment scheme for a fictitious railroad.

| | Insp./Foreman | | | | Maintainter/Asst. Insp. | | | | | | | M.T. | Elec. Tech. | | | |
|---|---|---|---|---|---|---|---|---|---|---|---|---|---|---|---|---|
| | Shift | | | | Shift Allocation | | | | | | | | Shift | | | |
| | Alloc. | | | | A. Insp. | | | Maint. | | | | Shift | Alloc. | | | |
| **Tower** | **1** | **2** | **3** | **Total** | **1** | **2** | **3** | **1** | **2** | **3** | **Total** | **3** | **1** | **2** | **3** | **Total** |
| SH | | | | | 1 | | | 2 | 3 | | 6 | | | | | |
| PK | | | | | 1 | | | 1 | 2 | 2 | 6 | | | | | |
| ST | | | | | 1 | | | 2 | 3 | 2 | 8 | | | | | |
| SD | 2 | 1 | 1 | 4 | 3 | | | 10 | 7 | 2 | 22 | | 3 | 2 | 3 | 8 |
| WK | | | | | 1 | | | 3 | 3 | 3 | 10 | | | | | |
| 43 | | | | | 1 | | | 1 | 2 | | 4 | | | | | |
| CN | 1 | 1 | | 2 | 1 | | | 1 | 2 | 2 | 6 | | 2 | 2 | | 4 |
| DE | | | | | 1 | | | 1 | 2 | 2 | 6 | | | | | |
| NH | | | | | 1 | | | 2 | 3 | 2 | 8 | 4 | | | | |
| **East** | **3** | **2** | **1** | **6** | **11** | **0** | **0** | **23** | **27** | **15** | **76** | **4** | **5** | **4** | **3** | **12** |
| TA | 1 | | | 1 | 1 | 1 | 1 | 1 | 1 | 1 | 6 | | | | | |
| TU | | | | | 1 | | | 1 | 2 | | 4 | | | | | |
| NK | | | | | 1 | | | 1 | | | 2 | | | | | |
| MO | 1 | 1 | 1 | 3 | 3 | | | 7 | 5 | 2 | 17 | 4 | 2 | 1 | 2 | 5 |
| WL | | | | | 1 | | | 2 | 3 | | 6 | | | | | |
| NW | | | | | 1 | | | 1 | 2 | 2 | 6 | | | | | |
| **South** | **2** | **1** | **1** | **4** | **8** | **1** | **1** | **13** | **13** | **5** | **41** | **4** | **2** | **1** | **2** | **5** |
| DV | | | | | 1 | | | 2 | 3 | | 6 | | | | | |
| HA | | | | | | | | | | | | | 1 | 1 | | 2 |
| OW | | | | | 1 | | | 1 | 2 | | 4 | | | | | |
| HM | 1 | | | 1 | 3 | | | 9 | 7 | 3 | 22 | | | | | |
| PS | 1 | 1 | 1 | 3 | | | | | | | | 5 | | | | |
| PK | | | | | 1 | | | 2 | 3 | 3 | 9 | | 1 | 1 | | 2 |
| CA | 1 | 1 | | 2 | 1 | | | 1 | 2 | 2 | 6 | | 1 | 1 | | 2 |
| HR | | | | | 1 | | | 1 | 2 | 2 | 6 | | | | | |
| CH | | | | | 1 | | | 2 | 3 | 2 | 8 | | | | | |
| BR | 1 | 1 | | 2 | 1 | | | 2 | 3 | 3 | 9 | | | | | |
| ET | | | | | | | | | | | | | 2 | 2 | 2 | 6 |
| **North** | **4** | **3** | **1** | **8** | **10** | **0** | **0** | **20** | **25** | **15** | **70** | **5** | **5** | **5** | **2** | **12** |
| ***System Total*** | **9** | **6** | **3** | ***18*** | **29** | **1** | **1** | **56** | **65** | **35** | ***187*** | ***13*** | **12** | **10** | **7** | ***29*** |

**Figure 12.** Personnel Assignment by Craft, Location, Gang, and Shift for a Sample System

This data could be utilized in a number of ways. Comparing these theoretical assignments with the actual assignments being utilized operationally today would provide useful information (and analytical rationale) on whether positions need to be established or abolished at certain locations on certain shifts. If total number of positions required (by craft) does not match the currently authorized positions per the corporate budget, the Signal maintenance dept. could make a representation to the corporate budget area, utilizing this data as justification to add, transfer, or reduce positions as necessary.

One of the recurring headaches experienced by maintenance managers is the constant demands by financial executives and outside stakeholders to justify overtime expenditure after-the-fact. The normal strategy is to cite dramatic non-recurring events or high profile failures, but this data actually provides an analytical basis to quantitatively indicate whether the maintenance organization is already under a level of staffing stress. If current employees on the payroll (by craft) is less than the theoretically required number to cover the workload, then extraordinary operational measures (such as assigning overtime—or pushing a larger-than-usual inspection quota on remaining staff) are already having to be taken to keep up with normal maintenance workload. This reduces the organization's capability to absorb sudden short-term increases in workload (e.g. Positive Train Control related signal system modifications), and practically guarantees a baseline level of overtime usage even when everything is going well.

***Alternate Work Assignment Scenario***

Signal Maintainers are Federally limited by hours of service regulations to a maximum of 12-hour shifts. In theory, two 12-hour shifts could cover the entire day, however, this would leave absolutely no surge capacity to catch up on backlog or absorb additional work, and will have long-term fatigue impacts [4]. It is also theoretically possible to assign shifts that cover off-peak hours only and therefore could be far more productive with preventative maintenance, but doing so would leave rush-hours uncovered except by dedicated trouble ticket teams, resulting in longer response times and train delays.

We understood from long-serving maintenance managers that at one point in the past, one railroad had a similar arrangement, that the overnight shift were not assigned except at one strategic locations per district with "trouble trucks" that would cover a wide geographic area. Paradoxically, the driver for this arrangement was employee preference, in that signalmen preferred not to make tests in darkness. This strategy of minimal overnight staffing led to a number of undesirable phenomena: (1) inevitably, immediately prior to the morning rush hour, trouble arose in multiple location and coverage could not be found to attend to them simultaneously; (2) because "trouble trucks" covered a large area, the maintainers are less familiar with the circuits they were working on, and did not have hands-on knowledge of recent maintenance history at that location, as a result took longer to find faults; (3) because overnight shifts are undesirable, and troubleshooting assignments are complex, due to the nature of the bid and bump process, these assignments inevitably ended up with the least experienced employees, who found themselves in the unenviable position of having to protect the rush-hour with very little support. In theory, it may be possible to create a job classification of "troubleshooting signalmen" and require a higher level of training and skills, however, this still could not provide the intimate knowledge of the plant that comes from making tests on a daily basis within their area, and anyway from a management practice point of view, it is important to promote "ownership" of the area that maintainers are responsible for, much like "section gangs" of yesteryear.

For these reasons we felt that segregation of maintenance and troubleshooting tasks were not really workable in practice.

Although the cost impacts of these scenarios could be tested with this model as part of workforce optimization efforts, it is outside of the scope of the current model to simulate the impact on component reliability or troubleshooting effectiveness as a result of changes to assignment patterns and task grouping strategy. We therefore did not test these scenarios because we felt the non-workforce related consequences of these methods of operation were sufficiently undesirable as to be not worth pursuing. Nonetheless, the assignment costing capability of this model could allow us to quantify the true cost of this manning strategy which protects the rush hour and promotes location-based accountability.

Discussion of these assignment strategies highlight the fact that maintenance assignments are not a simple mathematical question of providing the necessary coverage at the lowest cost. Substantial expertise in the nature of the tasks is needed to make informed decisions about signal assignment changes, and these decisions can have important reliability and safety impacts, sometimes in a difficult-to-quantify way.

## SUMMARY AND CONCLUSIONS

We described an asset management approach in a railway environment by partnering with maintenance managers (who have practical and hard-to-duplicate knowledge of asset condition and day-to-day maintenance needs) and began by inventorying maintenance employee positions and their daily tasks, rather than inventorying physical assets. This approach eventually produces a list of assets (and maintenance burdens each represented) organized logically, which could be used within a general asset management framework to inform strategic and investment decisions. However, prior to realizing that goal, this approach produces immediate benefits by vastly improving the understanding of maintenance activities, quantitatively assessing maintenance needs, and providing outputs that inform maintenance personnel assignment and rightsizing in the short term. These outputs from the "maintenance resourcing" model allows maintenance managers to precisely articulate resourcing needs (of when, where, what) and also allows financial managers to redistribute or reduce headcounts that could not be justified by the workload. Additionally, the data can be used to generate metrics that indicate to what extent each constraint on maintenance assignments decreases productivity, and extent to which maintenance burden could be realistically reduced through more aggressive failure reduction.

## Acknowledgement

The authors gratefully acknowledge the contributions of the following members of the Metro-North family in research assistance, education, mentoring, and making this work possible: J. Deptulski, A. Marussich, M. Bunnell, S. Hudson, R. Cervini, R. Wright, N. Colon, H. Farquharson, M. Najam-Ud-Din, S. Patel, A. Barron, F. Gimenez, S. Kennaway, E. Alcantara, Y. Schwartzman, J. Donnelly, M. Dabrowski, D. Altro, R. Culhane, D. Melillo, and J.E. Kesich. We'd also like to acknowledge the contribution of a number of external reviewers who wishes to remain anonymous. Responsibility for errors or omissions remains with the authors. All opinions expressed or implied are the authors' and do not necessarily reflect official policy or positions of the New York State Metropolitan Transportation Authority.